\RequirePackage{fix-cm} 
\documentclass[a4paper, twoside, reqno, dvips, 12pt]{amsart}
\usepackage{fixltx2e}   

\usepackage{etex}

\usepackage[latin1]{inputenc}
\usepackage[T1]{fontenc}

\usepackage{esint}
\usepackage{dsfont}
\usepackage{xspace}
\usepackage{amsgen}
\usepackage{amsthm}
\usepackage{amssymb}
\usepackage{amsmath}
\usepackage{wasysym}
\usepackage{upgreek}
\usepackage{amsfonts}
\usepackage{stmaryrd}
\usepackage{mathtools}

\usepackage{relsize}
\usepackage[mathcal, mathscr]{euscript}

\usepackage{mathrsfs}
\DeclareMathAlphabet{\mathscrbf}{OMS}{mdugm}{b}{n}

\usepackage{a4wide}

\usepackage[dvipsnames, table]{xcolor}
\definecolor{bckg}{RGB}{20.8, 20.8, 20.8}
\definecolor{oneblue}{rgb}{0.0, 0.0, 0.85}
\definecolor{Lightblue}{RGB}{214, 214, 214}
\definecolor{bluepigment}{rgb}{0.2, 0.2, 0.6}
\definecolor{charcoal}{rgb}{0.21, 0.27, 0.31}
\definecolor{denimblue}{rgb}{0.08, 0.38, 0.74}
\definecolor{Lightgray}{rgb}{0.89, 0.89, 0.89}
\definecolor{darkgrey}{rgb}{0.273, 0.281, 0.30}
\definecolor{darkelectricblue}{rgb}{0.33, 0.41, 0.47}

\usepackage[sort&compress, comma, square, numbers]{natbib}

\usepackage{psfrag}
\usepackage{graphicx}
\usepackage{subfigure}
\usepackage{morefloats}
\usepackage{indentfirst}

\usepackage{acronym}
\usepackage{microtype}
\usepackage[labelsep=period,%
            labelfont={bf,sf,color=bluepigment},%
            justification=raggedright]{caption}

\usepackage[perpage, symbol]{footmisc}

\usepackage[usenames, dvipsnames, pdf]{pstricks}
\usepackage{epsfig}
\usepackage{pst-grad} 
\usepackage{pst-plot} 

\usepackage[colorlinks,
           urlcolor=oneblue,
           linkcolor=denimblue,
           citecolor=NavyBlue,
           bookmarksopen=false,
           pdfpagemode=UseNone,
           pagebackref]{hyperref}

\usepackage[explicit]{titlesec}

\titleformat{\section}[block]
  {\color{NavyBlue}\Large\sffamily\bfseries}
  {}
  {0.0em}
  {\colorbox{bckg!5}{\strut\parbox{\dimexpr\linewidth-2\fboxsep\relax}{\thesection. #1}}}
  [\vspace*{0.33em}]

\titleformat{name=\section,numberless}[block]
  {\color{NavyBlue}\Large\sffamily\bfseries}
  {}
  {0.0em}
  {\colorbox{bckg!5}{\strut\parbox{\dimexpr\linewidth-2\fboxsep\relax}{#1}}}
  [\vspace*{0.33em}]

\titleformat{\subsection}
  {\color{NavyBlue}\large\sffamily\bfseries}
  {}
  {0.0em}
  {\colorbox{bckg!5}{\parbox{\dimexpr\linewidth-2\fboxsep\relax}{\thesubsection. #1}}}
  [\vspace*{0.33em}]

\titleformat{name=\subsection,numberless}
  {\color{NavyBlue}\Large\sffamily\bfseries}
  {}
  {0em}
  {\colorbox{bckg!5}{\parbox{\dimexpr\linewidth-2\fboxsep\relax}{#1}}}
  [\vspace*{0.33em}]

\titleformat{\subsubsection}
  {\color{bluepigment}\sffamily\normalsize\bfseries}
  {\thesubsubsection}
  {0.5em}
  {#1}
  [\vspace*{0.33em}]

\titleformat{\paragraph}[runin]
  {\color{bluepigment}\sffamily\small\bfseries}
  {}
  {0em}
  {#1}

\titlespacing{\section}{0.0em}{1.5em plus 2pt minus 2pt}%
{1.0em plus 2pt minus 2pt}[0em]
\titlespacing{\subsection}{0.5em}{1.5em plus 2pt minus 2pt}%
{1.0em}[0em]
\titlespacing{\subsubsection}{0.5em}{1.5em plus 2pt minus 2pt}%
{1.0em plus 2pt minus 2pt}[0em]

\usepackage{titletoc}

\contentsmargin{0.5em}
\newlength{\tocsep} 
\titlecontents{section}[\tocsep]
  {\addvspace{10pt}\bfseries\sffamily}
  {\contentslabel[\thecontentslabel]{\tocsep}}
  {}
  {\ \titlerule*[0.75pc]{.}\ \thecontentspage}
  []
\titlecontents{subsection}[\tocsep]
  {\addvspace{8pt}\sffamily}
  {\contentslabel[\thecontentslabel]{\tocsep}}
  {}
  {\ \titlerule*[0.5pc]{.}\ \thecontentspage}
  []
\titlecontents*{subsubsection}[\tocsep]
  {\addvspace{2pt}\footnotesize\sffamily}
  {}
  {}
  {\ \titlerule*[0.35pc]{.}\ \thecontentspage}
  [\\*]

\makeatletter
\def\@setauthors{%
  \begingroup
  \def\thanks{\protect\thanks@warning}%
  \trivlist
  \centering\footnotesize \@topsep30\p@\relax
  \advance\@topsep by -\baselineskip
  \item\relax
  \author@andify\authors
  \def\\{\protect\linebreak}%
  \textsc{\normalsize\textcolor{darkelectricblue}{\authors}}%
  \ifx\@empty\contribs
  \else
    ,\penalty-3 \space \@setcontribs
    \@closetoccontribs
  \fi
  \endtrivlist
  \endgroup
}
\def\@settitle{\begin{center}%
  \baselineskip14\p@\relax
    \bfseries
    \textsc{\Large\textcolor{charcoal}{\@title}}
  \end{center}%
}
\makeatother

\usepackage{enumitem}
\setlist[description]{%
  topsep=30pt,               
  itemsep=5pt,               
  font={\bfseries\sffamily\color{NavyBlue}}, 
}

\usepackage{fancyhdr}
\usepackage{lastpage}

\newcommand*\Title{\textcolor{bluepigment}{Dispersive shallow water wave modelling. Part I}}
\newcommand*\Authors{\textcolor{bluepigment}{G.~Khakimzyanov, D.~Dutykh, et al.}}
\newcommand*{\plogo}{\textcolor{gray}{{\texttt{arXiv.org} / \textsc{hal}}}} 

\numberwithin{equation}{section}

\newtheorem{remark}{Remark}

\newcommand{\up}[1]{$^{\mathrm{\small\textsf{#1}}}$} 

\newcommand{\ub}{\bar{\u}}
\newcommand{\ubar}{\bar{u}}
\newcommand{\pc}{\check{p}}

\newcommand{\R}{\mathds{R}}

\newcommand{\ut}{\tilde{\u}}

\newcommand{\Id}{\mathbb{I}}
\newcommand{\A}{\mathscr{A}}
\newcommand{\B}{\mathscr{B}}
\newcommand{\J}{\mathscr{U}}

\newcommand{\ud}{\mathrm{d}}

\newcommand{\E}{\mathscr{E}}

\newcommand{\Y}{\mathcal{Y}}

\newcommand{\Cs}{\mathscr{C}}

\newcommand{\Ru}{\mathcal{R}}

\newcommand{\Pp}{\mathscr{P}}
\newcommand{\Rr}{\mathscr{R}}
\renewcommand{\beta}{\upbeta}

\renewcommand{\leq}{\leqslant}
\renewcommand{\geq}{\geqslant}
\newcommand{\eps}{\varepsilon}
\renewcommand{\O}{\mathcal{O}}

\renewcommand{\H}{\mathcal{H}}

\newcommand{\D}{\mathscrbf{D}}
\renewcommand{\alpha}{\upalpha}

\newcommand{\x}{\boldsymbol{x}}
\newcommand{\U}{\boldsymbol{U}}
\newcommand{\vO}{\boldsymbol{0}}

\renewcommand{\u}{\boldsymbol{u}}
\renewcommand{\v}{\boldsymbol{v}}
\newcommand{\const}{\mathrm{const}}
\newcommand{\phic}{\check{\varphi}}

\newcommand{\omb}{\boldsymbol{\omega}}
\newcommand{\Cc}{\boldsymbol{\mathrm{C}}}

\newcommand{\Su}{\mathsf{S_U}}

\newcommand{\ie}{\emph{i.e.}\/ }
\newcommand{\eg}{\emph{e.g.}\/ }
\newcommand{\etc}{\emph{etc.}\/}
\newcommand{\etal}{\emph{et al.}\/ }

\renewcommand{\sim}{\thicksim}
\renewcommand{\div}{\grad\scal}

\newcommand{\Mat}{\mathrm{Mat}\,}
\newcommand{\scal}{\boldsymbol{\cdot}}
\newcommand{\grad}{\boldsymbol{\nabla}}
\newcommand{\timesb}{\boldsymbol{\times}}
\newcommand{\abs}[1]{\lvert\, #1\, \rvert}

\newcommand{\norm}[1]{\lVert\, #1\, \rVert}
\newcommand{\otimesb}{\boldsymbol{\otimes}}

\newcommand{\eqdef}{\mathop{\stackrel{\mathrm{def}}{:=}}}
\newcommand{\defeq}{\mathop{\stackrel{\mathrm{def}}{=:}}}
\newcommand{\od}[2]{\frac{\mathrm{d} #1}{\mathrm{d}\/#2}}

\newcommand{\half}{{\textstyle{1\over2}}}
\newcommand{\third}{{\textstyle{1\over3}}}

\newcommand{\sixth}{{\textstyle{1\over6}}}

\usepackage{acronym}
\acrodef{bvp}[BVP]{Boundary Value Problem}
\acrodef{NSWE}{Nonlinear Shallow Water Equations}

\begin{document}

\title[\Title]{Dispersive shallow water wave modelling. Part I: Model derivation on a globally flat space}

\author[G.~Khakimzyanov]{Gayaz Khakimzyanov}
\address{\textbf{G.~Khakimzyanov:} Institute of Computational Technologies, Siberian Branch of the Russian Academy of Sciences, Novosibirsk 630090, Russia}
\email{Khak@ict.nsc.ru}

\author[D.~Dutykh]{Denys Dutykh$^*$}
\address{\textbf{D.~Dutykh:} LAMA, UMR 5127 CNRS, Universit\'e Savoie Mont Blanc, Campus Scientifique, F-73376 Le Bourget-du-Lac Cedex, France}
\email{Denys.Dutykh@univ-savoie.fr}
\urladdr{http://www.denys-dutykh.com/}
\thanks{$^*$ Corresponding author}

\author[Z.~I.~Fedotova]{Zinaida Fedotova}
\address{\textbf{Z.~I.~Fedotova:} Institute of Computational Technologies, Siberian Branch of the Russian Academy of Sciences, Novosibirsk 630090, Russia}
\email{zf@ict.nsc.ru}

\author[D.~Mitsotakis]{Dimitrios Mitsotakis}
\address{\textbf{D.~Mitsotakis:} Victoria University of Wellington, School of Mathematics, Statistics and Operations Research, PO Box 600, Wellington 6140, New Zealand}
\email{dmitsot@gmail.com}
\urladdr{http://dmitsot.googlepages.com/}



\begin{titlepage}
\thispagestyle{empty} 
\noindent
{\Large Gayaz \textsc{Khakimzyanov}}\\
{\it\textcolor{gray}{Institute of Computational Technologies, Novosibirsk, Russia}}
\\[0.02\textheight]
{\Large Denys \textsc{Dutykh}}\\
{\it\textcolor{gray}{CNRS--LAMA, Universit\'e Savoie Mont Blanc, France}}
\\[0.02\textheight]
{\Large Zinaida \textsc{Fedotova}}\\
{\it\textcolor{gray}{Institute of Computational Technologies, Novosibirsk, Russia}}
\\[0.02\textheight]
{\Large Dimitrios \textsc{Mitsotakis}}\\
{\it\textcolor{gray}{Victoria University of Wellington, New Zealand}}
\\[0.08\textheight]

\vspace*{1cm}

\colorbox{Lightblue}{
  \parbox[t]{1.0\textwidth}{
    \centering\huge\sc
    \vspace*{0.7cm}
    
    \textcolor{bluepigment}{Dispersive shallow water wave modelling. Part I: Model derivation on a globally flat space}
    
    \vspace*{0.7cm}
  }
}

\vfill 

\raggedleft     
{\large \plogo} 
\end{titlepage}


\newpage
\thispagestyle{empty} 
\par\vspace*{\fill}   
\begin{flushright} 
{\textcolor{denimblue}{\textsc{Last modified:}} \today}
\end{flushright}


\newpage
\maketitle
\thispagestyle{empty}


\begin{abstract}

In this paper we review the history and current state-of-the-art in modelling of long nonlinear dispersive waves. For the sake of conciseness of this review we omit the unidirectional models and focus especially on some classical and improved \textsc{Boussinesq}-type and \textsc{Serre}--\textsc{Green}--\textsc{Naghdi} equations. Finally, we propose also a unified modelling framework which incorporates several well-known and some less known dispersive wave models. The present manuscript is the first part of a series of two papers. The second part will be devoted to the numerical discretization of a practically important model on moving adaptive grids.


\bigskip\bigskip
\noindent \textbf{\keywordsname:} long wave approximation; nonlinear dispersive waves; shallow water equations; solitary waves \\


\end{abstract}


\newpage
\tableofcontents
\thispagestyle{empty}


\newpage
\section{Introduction}

The history of nonlinear dispersive modelling goes back to the end of the XIX\up{th} century \cite{Craik2004}. At that time J.~\textsc{Boussinesq} (1877) \cite{Boussinesq1877} proposed (in a footnote on page 360) the celebrated \textsc{Korteweg}--\textsc{de Vries} equation, re-derived later by D.~\textsc{Korteweg} \& G.~\textsc{de Vries} (1895) \cite{KdV}. Of course, J.~\textsc{Boussinesq} proposed also the first \textsc{Boussinesq}-type equation \cite{Boussinesq1871, Boussinesq1872} as a theoretical explanation of \emph{solitary waves} observed earlier by J.~\textsc{Russell} (1845) \cite{Russell1845}. After this initial active period there was a break in this field until 1950's. The silence was interrupted by the new generation of `pioneers' --- F.~\textsc{Serre} (1953) \cite{Serre1953, Serre1953a}, C.C.~\textsc{Mei} \& \textsc{Le M\'ehaut\'e} (1966) \cite{Mei1966} and D.~\textsc{Peregrine} (1967) \cite{Peregrine1967} who derived modern nonlinear dispersive wave models. After this time the modern period started, which can be characterized by the proliferation of journal publications and it is much more difficult to keep track of these records. Subsequent developments can be conventionally divided in two classes:
\begin{enumerate}
  \item Application and critical analysis of existing models in new (and often more complex) situations
  \item Development of new high-fidelity physical approximate models
\end{enumerate}
Sometimes both points can be improved in the same publication. We would like to mention that according to our knowledge the first applications of \textsc{Peregrine}'s model \cite{Peregrine1967} to three-dimensional practical problems were reported in \cite{Abbott1978, Rygg1988}.

In parallel, scalar model equations have been developed. They describe the unidirectional wave propagation \cite{Olver1988a, Dutykh2010e}. For instance, after the above-mentioned KdV equation, its regularized version was proposed first by \textsc{Peregrine} (1966) \cite{Peregrine1966}, then by \textsc{Benjamin}, \textsc{Bona} \& \textsc{Mahony} (1972) \cite{bona}. Now this equation is referred to as the Regularized Long Wave (RLW) or \textsc{Benjamin}--\textsc{Bona}--\textsc{Mahony} (BBM) equation. In \cite{bona} the well-posedness of RLW/BBM equation in the sense of J.~\textsc{Hadamard} was proven as well. Even earlier \textsc{Whitham} (1967) \cite{Whitham1967a} proposed a model equation which possesses the dispersion relation of the full \textsc{Euler} equations (it was constructed in an ad-hoc manner to possess this property). It turned out to be an excellent approximation to the \textsc{Euler} equations in certain regimes \cite{Moldabayev2015}. Between unidirectional and bi-directional models there is an intermediate level of scalar equations with second order derivatives in time. Such an intermediate model was proposed, for example, in \cite{Kim1988}. Historically, the first \textsc{Boussinesq}-type equation proposed by J.~\textsc{Boussinesq} \cite{Boussinesq1877} was in this form as well. The main advantage of these models is their simplicity on one hand, and the ability of providing good quantitative predictions on the other hand.

One possible classification of existing nonlinear dispersive wave models can be made upon the choice of the horizontal velocity variable. Two popular choices were suggested in \cite{Peregrine1967}. Namely, one can use the depth-averaged velocity variable (see \eg \cite{Wu1981, Zheleznyak1985, Rygg1988, Dellar2005, Fedotova2009, Fedotova2014}). Usually, such models enjoy nice mathematical properties such as the exact mass conservation equation. The second choice consists in taking the trace of the velocity on a surface defined in the fluid bulk $y\ =\ \Y(\x,\,t)$. Notice, that surface $\Y(\x,\,t)$ may eventually coincide with the free surface \cite{Craig1994} or with the bottom \cite{Aleshkov1996, Mei1966}. This technique was used for the derivation of several Boussinesq type systems with flat bottom, initially in \cite{BS} and later in \cite{BC, BCS} and analysed thoroughly theoretically and numerically in \cite{BC, Bona2004, ADM1, ADM2, Antonopoulos2010, DMII}. Sometime the choice of the surface is made in order to obtain a model with improved dispersion characteristics \cite{BS, Madsen1991, Wei1995}. One of the most popular model of this class is due to O.~\textsc{Nwogu} (1993) \cite{Nwogu1993} who proposed to use the horizontal velocity defined at $y\ =\ \Y(\x)\ \eqdef\ \beta\, h(\x)$ with $\beta\ \approx\ 0.531$. This result was improved in \cite{Simarro2013a} to $\beta\ \approx\ 0.555$ (taking into consideration the shoaling effects as well). However, it was shown later that this theoretical `improvement' is immaterial when it comes to the description of real sea states \cite{Choi2015}.

Later, other choices of surface $\Y(\x,\,t)$ were proposed. For example, in \cite{Kennedy2001, Lynett2002} the surface $\Y(\x,\,t)$ was chosen to be genuinely unsteady (due to the free surface and/or bottom motion). This choice was motivated by improving also the nonlinear characteristics of the model. Some other attempts can be found in \cite{Wei1995, Hsiao2002, LWL, Chen2006a, Mitsotakis2007}. On the good side of these models we can mention accurate approximation of the dispersion relation up to intermediate depths and in some cases good well-posedness results. On the other side, equations are often cumbersome with unclear mathematical properties (\eg well-posedness, existence of travelling waves, \etc). Below we shall discuss more closely some of the models of this type.

For another recent complementary review of \textsc{Boussinesq}-type and other nonlinear dispersive models, which discusses also applications and some numerical approaches we refer to \cite{Brocchini2013} and for a detailed analysis of the theory and asymptotics for the water-wave problem we refer to \cite{Lannes2013}.

This manuscript is the first part in a series of four papers (the other parts are \cite{Khakimzyanov2016, Khakimzyanov2016a, Khakimzyanov2016b}). Here we attempt to make a literature review on the topic of nonlinear weakly dispersive wave modelling in shallow water environments. This topic is so broad that we apologize in advance if we forgot to mention someone's work. It was not made on purpose. Moreover, we propose a unified modeling framework which encompasses some more or less known models in this field. Namely, we show how several well-known models can be derived from the base model by making judicious choices of dynamic variables and/or their fluxes. We also try to point out some important properties of some model equations that have not attracted so much the attention of the researchers. The second part will be devoted to some numerical questions \cite{Khakimzyanov2016}. More precisely, we shall propose an adaptive finite volume discretization of a particular widely used dispersive wave model. The numerical method adaptivity is achieved by moving grid points to the locations where it is needed. The title of the first two parts include the wording `on a globally flat space'. It means basically that we consider a fluid flow with free surface on a Cartesian space, even if some bathymetry variations\footnote{The amount of bathymetry variations allowed in our modelling will be discussed in the second part \cite{Khakimzyanov2016} of this series.} are allowed, \ie the bottom is not necessarily flat. The (globally) spherical geometries will be discussed in some detail in Parts~\textsc{III} \& \textsc{IV} \cite{Khakimzyanov2016a, Khakimzyanov2016b}.

The present article is organized as follows: In Section~\ref{sec:derive} we derive the base model. However, the derivation procedure is quite general and it can be used to derive many other particular models, some of them being well-known and some possibly new. In Section~\ref{sec:weak} we propose also a weakly nonlinear version of the base model. Finally, in Section~\ref{sec:concl} we outline the main conclusions and perspectives of the present study.


\section{Base model derivation}
\label{sec:derive}

First of all we describe the physical problem formulation along with underlying constitutive assumptions. Later on this formulation will be further simplified using the asymptotic (or perturbation) expansions methods \cite{Nayfeh2000}.

Consider the flow of an ideal incompressible liquid in a physical three-dimensional space. We assume additionally that the fluid is homogeneous (\ie~the density $\rho\ =\ \const$) and the gravity acceleration $g$ is constant everywhere\footnote{This assumption is quite realistic since the variation of this parameter around the Earth is less than 1\%.}. Without any loss of generality from now on we can set $\rho\ \equiv\ 1$. For the sake of simplicity, in this study we neglect all other forces (such as the \textsc{Coriolis} force and friction). Hence, we deal with pure gravity waves.

\begin{figure}
  \centering
  \includegraphics[width=0.75\textwidth]{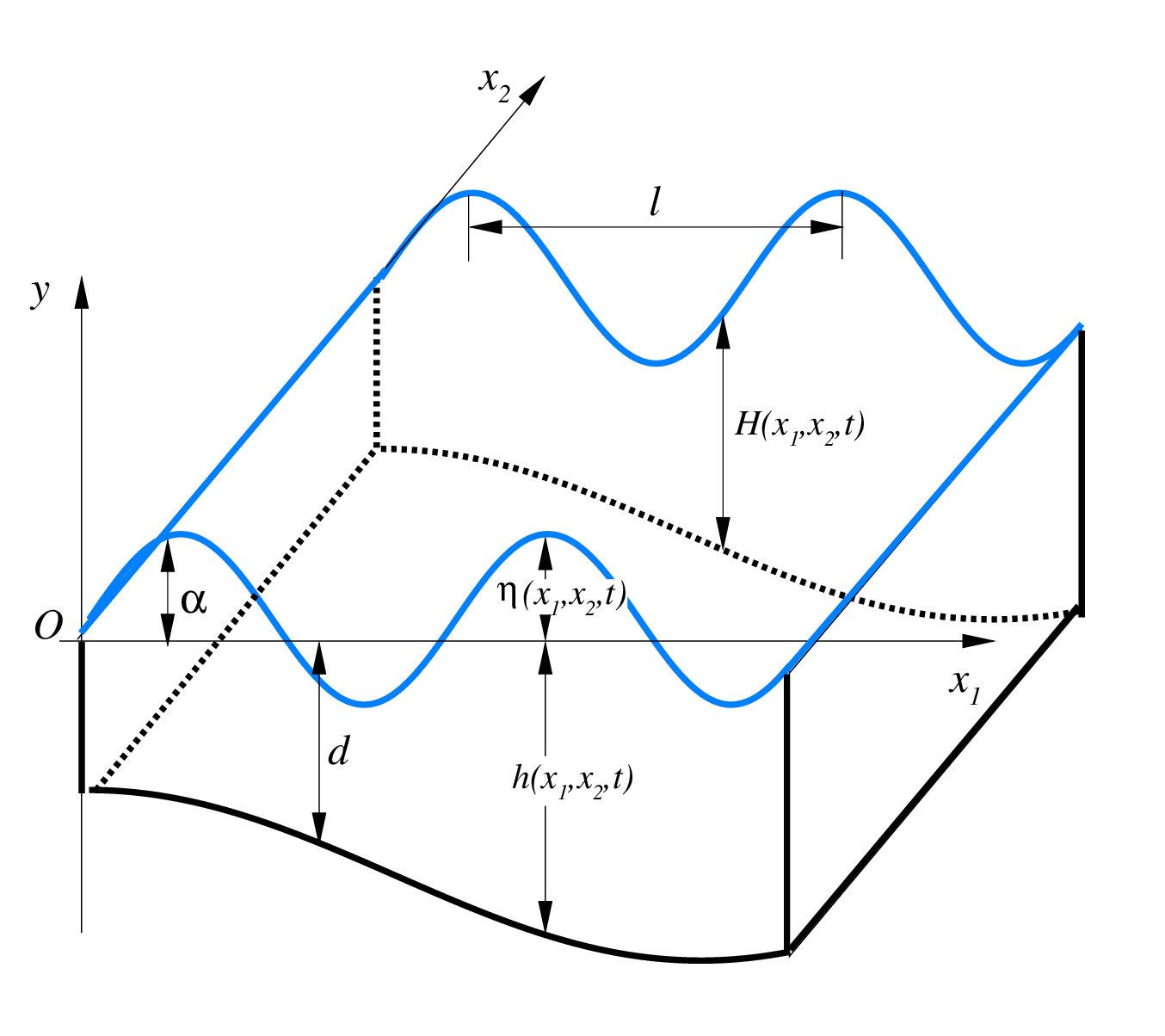}
  \caption{\small\em Sketch of the fluid domain.}
  \label{fig:sketch}
\end{figure}

In order to describe the mathematical model, we introduce a Cartesian coordinate system $O x_1 x_2\, y$. The horizontal plane $O x_1 x_2$ coincides with the still water level $y\ =\ 0$ and the axis $O y$ points vertically upwards. By vector $\x\ =\ (x_1,\,x_2)$ we denote the horizontal coordinates. The fluid layer is bounded below by the solid (impenetrable) bottom $y\ =\ -h\,(\x,\,t)$ and above by the free surface $y\ =\ \eta\,(\x,\,t)$. The sketch of the fluid domain is schematically shown in Figure~\ref{fig:sketch}.

The flow is considered to be completely determined if we find the velocity field $\U(\x,\,y,\,t)\ =\ \bigl(\u(\x,\,y,\,t),\,v(\x,\,y,\,t)\bigr)$ ($\u = (u_1,\,u_2)$ being the horizontal velocity components) along with the pressure field $p(\x,\,y,\,t)$ and the free surface elevation $\eta(\x,\,t)$, which satisfy the system of \textsc{Euler} equations:
\begin{align}\label{eq:euler1}
  \div\u\ +\ v_y\ &=\ 0\,, \\
  \u_t\ +\ (\u\scal\grad)\u\ +\ v\,\u_y\ +\ \grad p\ &=\ 0\,,\label{eq:euler2} \\
  v_t\ +\ \u\scal\grad v\ +\ v\, v_y\ +\ p_y\ &=\ -g\,,\label{eq:euler3}
\end{align}
where $\grad\ =\ (\partial_{x_1},\, \partial_{x_2})$ denotes the horizontal gradient operator. The \textsc{Euler} equations are completed with free surface kinematic and dynamic boundary conditions
\begin{align}\label{eq:bc1}
  \eta_t\ +\ \u\scal\grad\eta\ &=\ v\,, \quad y\ =\ \eta(\x,\,t)\,, \\
  p\ &=\ 0\,, \quad y\ =\ \eta(\x,\,t)\,. \label{eq:bc2}
\end{align}
Finally, on the bottom we impose the impermeability condition (\ie~the fluid particles cannot penetrate the solid boundary), which states that the normal velocity on the bottom vanishes:
\begin{equation}\label{eq:bc3}
  h_t\ +\ \u\scal\grad h\ +\ v\ =\ 0\,, \quad y\ =\ -h(\x,\,t)\,.
\end{equation}
Below we shall discuss also the components of the vorticity vector $\omb\ =\ \grad\timesb\ \U$, which are given by
\begin{align*}
  \omega_1\ &=\ v_{x_2}\ -\ u_{2,\,y}\,, \\
  \omega_2\ &=\ -v_{x_1}\ +\ u_{1,\,y}\,, \\
  \omega_3\ &=\ u_{2,\,x_1}\ -\ u_{1,\,x_2}\,.
\end{align*}

\subsection{Dimensionless variables}

In order to study the propagation of long gravity waves, we have to scale the governing equations \eqref{eq:euler1}--\eqref{eq:euler3} along with the boundary conditions \eqref{eq:bc1}--\eqref{eq:bc3}. For this purpose we choose characteristic scales of the flow. Let $\ell$, $d$ and $\alpha$ be the typical (wave or basin) length, water depth and wave amplitude correspondingly (they are depicted in Figure~\ref{fig:sketch}). Then, dimensionless independent variables can be introduced as follows
\begin{equation*}
  x_{1,2}^{\ast}\ =\ \frac{x_{1,2}}{\ell}\,, \quad
  y^{\ast}\ =\ \frac{y}{d}\,, \quad
  t^{\ast}\ =\ \frac{t}{\ell/\sqrt{g d}}\,.
\end{equation*}
The dependent variables are scaled\footnote{We would like to make a comment about the pressure scaling. For dimensional reasons we added in parentheses the fluid density $\rho$. However, it is not present in governing equations since for an incompressible flow of a homogeneous liquid $\rho$ can be set to the constant $1$ without loss of generality.} as
\begin{equation*}
  h^{\ast}\ =\ \frac{h}{d}\,, \quad
  \eta^{\ast}\ =\ \frac{\eta}{\alpha}\,, \quad
  p^{\ast}\ =\ \frac{p}{(\rho) g d}\,, \quad
  \u^{\ast}\ =\ \frac{\u}{\sqrt{gd}}\,, \quad
  v^{\ast}\ =\ \frac{v}{d\sqrt{gd}/\ell}\,.
\end{equation*}
The components of vorticity $\omb$ are scaled as
\begin{equation*}
  \omega_{1,2}^{\ast}\ =\ \frac{\omega_{1,2}}{\sqrt{gd}/d}\,, \qquad
  \omega_{3}^{\ast}\ =\ \frac{\omega_3}{\sqrt{gd}/\ell}\,.
\end{equation*}
The scaled version of the \textsc{Euler} equations \eqref{eq:euler1}--\eqref{eq:euler3} read now
\begin{align}\label{eq:euler4}
  \div\u\ +\ v_y\ &=\ 0\,, \\
  \u_t\ +\ (\u\scal\grad)\u\ +\ v\,\u_y\ +\ \grad p\ &=\ 0\,,\label{eq:euler5} \\
  \mu^2\,\bigl(v_t\ +\ \u\scal\grad v\ +\ v\, v_y\bigr)\ +\ p_y\ &=\ -1\,,\label{eq:euler6}
\end{align}
where we drop the asterisk symbol $\,\ast\,$ for the sake of notation compactness. Boundary conditions at the free surface similarly become
\begin{align}\label{eq:bc4}
  \eps\,\bigl(\eta_t\ +\ \u\scal\grad\eta\bigr)\ &=\ v\,, \quad y\ =\ \eps\,\eta(\x,\,t)\,, \\
  p\ &=\ 0\,, \quad y\ =\ \eps\,\eta(\x,\,t)\,. \label{eq:bc5}
\end{align}
It can be easily checked that the bottom boundary condition \eqref{eq:bc3} remains invariant under this scaling. Finally, the scaled components of vorticity $\omega^\ast$ are
\begin{align*}
  \omega_1\ &=\ \mu^2\,v_{x_2}\ -\ u_{2,\,y}\,, \\
  \omega_2\ &=\ -\mu^2\,v_{x_1}\ +\ u_{1,\,y}\,, \\
  \omega_3\ &=\ u_{2,\,x_1}\ -\ u_{1,\,x_2}\,.
\end{align*}

Above we introduced two important dimensionless parameters:
\begin{description}
  \item[Nonlinearity] $\eps\ \eqdef\ \dfrac{\alpha}{d}$, which measures the deviation of waves with respect to the unperturbed water level
  \item[Dispersion] $\mu\ \eqdef\ \dfrac{d}{\ell}$, which indicates how long the waves are comparing to the mean depth (or equivalently how shallow is the water)
\end{description}

\subsection{Long wave approximation}

In approximate shallow water systems the dynamic variables are the total water depth $\H(\x,\,t)\ \eqdef\ h(\x,\,t)\ +\ \eps\,\eta(\x,\,t)$ and some vector $\ub(\x,\,t)$ which is supposed to approximate the horizontal velocity vector of the full model $\u(\x,\,y,\,t)$. In many works $\ub(\x,\,t)$ is chosen as the trace of the horizontal velocity $\u$ at certain surface $y\ =\ \Y_\sigma(\x,\,t)$ in the fluid bulk \cite{Nwogu1993, Kennedy2001, Lynett2002}, \ie
\begin{equation}\label{eq:defY}
  \ub(\x,\,t)\ \eqdef\ \u\bigl(\x,\, \Y_\sigma(\x,\,t),\, t\bigr)\,.
\end{equation}
Another popular choice for the velocity variable consists in taking the depth-averaged velocity \cite{Serre1953, Peregrine1967, Rygg1988, Dellar2005, Dutykh2011a}:
\begin{equation}\label{eq:deptha}
  \ub(\x,\,t)\ \eqdef\ \frac{1}{\H(\x,\,t)}\;\int_{-h(\x,\,t)}^{\;\eps\,\eta(\x,\,t)}\, \u(\x,\,y,\,t)\;\ud y\,.
\end{equation}
By applying the mean value theorem \cite{zorich} to the last integral, we obtain that two approaches are mathematically formally equivalent:
\begin{equation*}
  \ub(\x,\,t)\ \equiv\ \u\bigl(\x,\, \Y_\xi(\x,\,t),\, t\bigr)\,.
\end{equation*}
However, this time the surface $y\ =\ \Y_\xi(\x,\,t)$ remains unknown, while above it was explicitly specified. We only know that such surface exists.

Below we shall consider only long wave approximation to the full Euler equations. Namely, we assume that $\ub(\x,\,t)$ approximates the true horizontal velocity $\u(\x,\,y,\,t)$ to the order $\O(\mu^{2})$, \ie
\begin{equation}\label{eq:hor}
  \u(\x,\,y,\,t)\ =\ \ub(\x,\,t)\ +\ \mu^{2}\,\ut(\x,\,y,\,t)\,.
\end{equation}
By integrating the continuity equation \eqref{eq:euler4} over the total depth and taking into account boundary conditions \eqref{eq:bc3}, \eqref{eq:bc4} we obtain the mass conservation equation
\begin{equation}\label{eq:mass}
  \H_t\ +\ \div(\H\,\ub)\ =\ -\mu^2\,\div(\H\J)\,,
\end{equation}
where
\begin{equation}\label{eq:defJ}
  \J(\x,\,t)\ \eqdef\ \frac{1}{\H(\x,\,t)}\,\int_{-h(\x,\,t)}^{\;\eps\eta\,(\x,\,t)}\, \ut(\x,\,y,\,t)\;\ud y\,.
\end{equation}
If we choose the variable $\ub$ to be depth-averaged, then $\J(\x,\,t)\ \equiv\ 0$ and the mass conservation equation \eqref{eq:mass} takes the very familiar form
\begin{equation*}
  \H_t\ +\ \div(\H\ub)\ =\ 0\,.
\end{equation*}
Integration of equation \eqref{eq:euler4} over the vertical coordinate in the limits from $-h(\x,\,t)$ to $y$ and taking into account the bottom boundary condition \eqref{eq:bc3} leads to the following representation for the vertical velocity in the fluid column:
\begin{equation}\label{eq:vert}
  v(\x,\,y,\,t)\ =\ -\D h\ -\ (y + h)\div\ub\ +\ \O(\mu^2)\,,
\end{equation}
where for the sake of simplicity we introduced the material (or total, or convective) derivative operator:
\begin{equation*}
  \D\,[\,\cdot\,]\ \eqdef\ [\,\cdot\,]_{\,t}\ +\ \ub\scal\grad\;[\,\cdot\,]\,.
\end{equation*}
Below the powers of this operator will appear in our computations:
\begin{equation*}
  \D^{k}\,[\,\cdot\,]\ \eqdef\ \underbrace{\D\cdot\D\cdot\ldots\cdot\D}_{k \mbox{ times}}\;[\,\cdot\,]\,, \qquad k\ \geq\ 1\,.
\end{equation*}

We have to express \emph{asymptotically} also the pressure field $p(\x,\,y,\,t)$ in terms of the dynamic variables $\bigl(\H(\x,\,t),\, \ub(\x,\,t)\bigr)$. Thus, we integrate the vertical momentum equation \eqref{eq:euler6} over the vertical coordinate in the limits from $y$ to the free surface:
\begin{equation}\label{eq:pint}
  p(\x,\,y,\,t)\ =\ \mu^2\,\int_{y}^{\;\eps\eta(\x,\,t)}\Bigl[\,\D\, v\ +\ v\,v_y\ +\ \O(\mu^2)\,\Bigr]\;\ud y\ -\ y\ +\ \eps\,\eta(\x,\,t)\,.
\end{equation}
The integrand can be expressed in term of $\H(\x,\,t)$ and $\ub(\x,\,t)$ using representation \eqref{eq:vert}:
\begin{equation*}
  \D\, v\ +\ v\,v_y\ =\ -(y + h)\,\Rr_1\ -\ \Rr_2\ +\ \O(\mu^2)\,,
\end{equation*}
where we defined
\begin{align*}
  \Rr_1(\x,\,t)\ &\eqdef\ \D\,\div\ub\ -\ (\div\ub)^2\,,\\
  \Rr_2(\x,\,t)\ &\eqdef\ \D^{2}h\,.
\end{align*}
Substituting the last result into the integral representation \eqref{eq:pint} and integrating it exactly in $y$ leads the following expression of the pressure field in the fluid layer:
\begin{equation}\label{eq:press}
  p\ =\ \H\ -\ (y + h)\ -\ \mu^2\,\biggl[\,\bigl(\H\ -\ (y + h)\bigr)\,\Rr_2\ +\ \Bigl(\frac{\H^{\,2}}{2}\ -\ \frac{(y + h)^2}{2}\Bigr)\,\Rr_1\,\biggr]\ +\ \O(\mu^4)\,.
\end{equation}
Notice that this representation does not depend on the expression of the velocity correction $\ut(\x,\,t)$. If in the last formula we neglect terms of $\O(\mu^4)$ and return to physical variables, we can obtain the pressure reconstruction formula in the fluid bulk:
\begin{equation*}
  \frac{p}{\rho}\ =\ g\,\Bigl[\,\H\ -\ (y + h)\,\Bigr]\ -\ \biggl[\,\bigl(\H\ -\ (y + h)\bigr)\,\Rr_2\ +\ \Bigl(\frac{\H^{\,2}}{2}\ -\ \frac{(y + h)^2}{2}\Bigr)\,\Rr_1\,\biggr]\,.
\end{equation*}
We underline the fact that the last formula is accurate to the order $\O(\mu^4)$. This formula will be used in \cite{Khakimzyanov2016} in order to reconstruct the pressure field under a solitary wave, which undergoes some nonlinear transformations.

In order to obtain an evolution equation for the approximate horizontal velocity $\ub(\x,\,t)$ we integrate over the vertical coordinate equation \eqref{eq:euler5}:
\begin{equation}\label{eq:intH}
  \int_{-h}^{\;\eps\eta}\bigl[\,\u_t\ +\ (\u\scal\grad)\u\ +\ v\,\u_y\,\bigr]\;\ud y\ +\ \grad\int_{-h}^{\;\eps\eta} p\;\ud y\ -\ \left.p\right|_{y = -h}\cdot\grad h\ =\ 0\,.
\end{equation}
The pressure variable can be easily eliminated from the last equation using the representation formula \eqref{eq:press}:
\begin{multline*}
  \grad\int_{-h}^{\;\eps\eta} p\;\ud y\ -\ \left.p\right|_{y = -h}\cdot\grad h\ =\\ =\ \eps\,\H \grad h\ -\ \mu^2\Bigl[\,\grad\bigl(\third \H^{\,3}\Rr_1\ +\ \half \H^{\,2}\Rr_2\bigr)\ -\ \H\,\grad h\,\bigl(\half \H\Rr_1\ +\ \Rr_2\bigr)\,\Bigr]\ +\ \O(\mu^4)\,.
\end{multline*}
Then, using the representation \eqref{eq:vert} for the vertical velocity $v$, we can write
\begin{align*}
  \int_{-h}^{\;\eps\eta} v\,\u_y\;\ud y\ &=\ -\mu^2\int_{-h}^{\;\eps\eta}\bigl[\,\D h\ +\ (y + h)\div\ub\,\bigr]\,\ut_y\;\ud y\ +\ \O(\mu^4) \\
  &=\ -\left.\mu^2(\D h)\cdot\ut\right|_{y\, =\, -h}^{y\, =\, \eps\eta}\ -\ \mu^2\,\div\ub\underbrace{\int_{-h}^{\;\eps\eta}(y + h)\ut_y\;\ud y}_{(\ast)}\ +\ \O(\mu^4)\,.
\end{align*}
The integral ($\ast$) can be computed using integration by parts
\begin{equation*}
  \int_{-h}^{\;\eps\eta}(y + h)\,\ut_y\;\ud y\ =\ \left.\H\cdot\ut\right|^{y\, =\, \eps\eta}\ -\ \H\J\,.
\end{equation*}
Combining together these results, we obtain the following asymptotic formula
\begin{equation*}
  \frac{1}{\mu^2}\int_{-h}^{\;\eps\eta} v\,\u_y\;\ud y\ =\ \left.(\D h)\cdot\ut\right|_{y\, =\, -h}\ -\ \bigl[\,\D h\ +\ \H\,\div\ub\,\bigr]\left.\ut\right|^{y\, =\, \eps\eta}\ +\ \H\J\div\ub\ +\ \O(\mu^2)\,.
\end{equation*}
Finally, we take care of convective terms
\begin{multline*}
  \int_{-h}^{\;\eps\eta}\bigl[\,\u_t\ +\ (\u\scal\grad)\u\,\bigr]\;\ud y\ =\ \int_{-h}^{\;\eps\eta}\D\ub\;\ud y\ +\ \mu^2\int_{-h}^{\;\eps\eta}\D\ut\;\ud y\ +\ \mu^2\int_{-h}^{\;\eps\eta}(\ut\scal\grad)\ub\;\ud y\ +\ \O(\mu^4)\\
  =\ \H\D\ub\ +\ \mu^2\Bigl[\,\D\bigl[\,\H\J\,\bigr]\ -\ \D[\,\eps\eta\,]\cdot\left.\ut\right|^{y\,=\,\eps\,\eta}\ -\ \D h\cdot\left.\ut\right|_{y\,=\,-h}\ +\ \H(\J\scal\grad)\,\ub\,\Bigr]\ +\ \O(\mu^4)\,.
\end{multline*}
Finally, we obtain
\begin{multline*}
  \int_{-h}^{\;\eps\eta}\bigl[\,\u_t\ +\ (\u\scal\grad)\u\ +\ v\,\u_y\,\bigr]\;\ud y\ =\ \H\D\ub\ -\ \mu^2\underbrace{\bigl[\,\D \H\ +\ \H\div\ub\,\bigr]}_{(\ast\ast)}\cdot\left.\ut\right|^{y\,=\,\eps\eta}\ + \\
  +\ \mu^2\,\Bigl[\,\D\bigl[\,\H\J\,\bigr]\ +\ \H(\J\scal\grad)\,\ub\ +\ \H\J\div\ub\,\Bigr]\,.
\end{multline*}
From the mass conservation equation \eqref{eq:mass} we have
\begin{equation*}
  \D \H\ +\ \H\,\div\ub\ =\ -\mu^2\,\div(\H\J)\ =\ \O(\mu^2)\,.
\end{equation*}
Thus, the term ($\ast\ast$) can be asymptotically neglected. As a result we have
\begin{equation*}
  \int_{-h}^{\;\eps\eta}\bigl[\,\u_t\ +\ (\u\scal\grad)\u\ +\ v\,\u_y\,\bigr]\;\ud y\ =\ \H\D\ub\ +\ \mu^2\,\Bigl[\,\D\bigl[\,\H\J\,\bigr]\ +\ \H(\J\scal\grad)\,\ub\ +\ \H\J\div\ub\,\Bigr]\ +\ \O(\mu^4)\,.
\end{equation*}
Substituting all these intermediate results into depth-integrated horizontal momentum equation \eqref{eq:intH}, we obtain the required evolution equation for $\ub$:
\begin{multline}\label{eq:ugly}
  \ub_t\ +\ (\ub\scal\grad)\ub\ +\ \eps\,\grad\eta\ =\ \frac{\mu^2}{\H}\;\Bigl[\,\grad\bigl(\third \H^{\,3}\Rr_1\ +\ \half \H^{\,2}\Rr_2\bigr)\ -\ \H\,\grad h\,\bigl(\half \H\Rr_1\ +\ \Rr_2\bigr)\,\Bigr]\\
  -\ \frac{\mu^2}{\H}\;\Bigl[\,\D\bigl[\,\H\J\,\bigr]\ +\ \H(\J\scal\grad)\,\ub\ +\ \H\J\div\ub\,\Bigr]\,.
\end{multline}
The last equation may look complicated. However, it can be rewritten in a clearer way by pointing out explicitly the non-hydrostatic pressure effects. It turns out that it is advantageous to introduce the depth-integrated (but not depth-averaged) pressure:
\begin{equation}\label{eq:press2}
  \Pp(\H,\,\ub)\ \eqdef\ \int_{-h}^{\;\eps\eta}\, p\;\ud y\ =\ \frac{\H^{\,2}}{2}\ -\ \mu^2\,\Bigl(\third \H^{\,3}\Rr_1\ +\ \half \H^{\,2}\Rr_2\Bigr)\,.
\end{equation}
We introduce also the pressure trace $\pc$ at the bottom:
\begin{equation*}
  \pc(x,\,t)\ \eqdef\ \left.p\right|_{y\,=\,-h}\ =\ \H\ -\ \mu^2\,\Bigl(\half \H^{\,2}\Rr_1\ +\ \H\Rr_2\Bigr)\,.
\end{equation*}
Using these new variables equation \eqref{eq:ugly} becomes
\begin{equation*}
  \ub_t\ +\ (\ub\scal\grad)\ub\ +\ \frac{\grad\Pp}{\H}\ =\ \frac{\pc\,\grad h}{\H}\ -\ \frac{\mu^2}{\H}\;\Bigl[\,(\H\J)_t\ +\ (\ub\scal\grad)(\H\J)\ +\ \H(\J\scal\grad)\,\ub\ +\ \H\J\div\ub\,\Bigr]\,.
\end{equation*}

The derived system of equations admits an elegant conservative form\footnote{This form becomes truly conservative (in the sense of hyperbolic conservation laws) only on the flat bottom, \ie~$h(\x,\,t)\ =\ h_0\ =\ \const\ \Rightarrow\ \grad h\ \equiv\ \vO\,$.}:
\begin{align}\label{eq:base1}
  \H_t\ +\ \div[\,\H\U\,]\ &=\ 0\,, \\
  (\H\,\U)_t\ +\ \div\Bigl[\,\H\ub\otimesb\U\ +\ \Pp(\H,\,\ub)\cdot\Id\ +\ \mu^2 \H\,\J\otimesb\ub\,\Bigr]\ &=\ \pc\,\grad h\,, \label{eq:base2}
\end{align}
where we introduced a new velocity variable $\U\ \eqdef\ \ub\ +\ \mu^2\,\J$ and $\Id\ \in\ \Mat_{2\,\times\,2}(\R)$ is the identity matrix. Operator $\otimesb$ is the tensorial product, \ie~for two vectors $\u\ \in\ \R^m$ and $\v\ \in\ \R^n$
\begin{equation*}
  \u\otimesb\v\ \eqdef\ (u_i\cdot v_j)^{1\, \leq\, i\, \leq\ m}_{1\, \leq\, j\, \leq\, n}\ \in\ \Mat_{m\,\times\,n}(\R)\,.
\end{equation*}
From now on equations \eqref{eq:base1}, \eqref{eq:base2} will be referred to as the \emph{base model} of our study. In order to close the last system of equations \eqref{eq:base1}, \eqref{eq:base2}, we have to express the variable $\J$ in terms of other dynamic variables $\H(\x,\,t)$ and $\ub(\x,\,t)$. Several popular choices will be discussed below. Notice also that \emph{nowhere} in the derivation above the flow irrotationality was assumed.

\begin{remark}
Notice that taking formally the limit $\mu\ \to\ 0$ in equations \eqref{eq:base1}, \eqref{eq:base2} yields straightforwardly the well-known Nonlinear Shallow Water (NSW or \textsc{Saint}-\textsc{Venant}) Equations \cite{SV1871}. Thus, our base model satisfies the \textsc{Bohr} \emph{correspondence principle}\footnote{This principle was formulated by Niels~\textsc{Bohr} (1920) \cite{Bohr1920}. Loosely speaking, this principle states that Quantum Mechanics reproduces Classical Mechanics in the limit of large quantum numbers. Correspondingly, a nonlinear dispersive model should describe correctly the propagation of non-dispersive waves in the limit when the dispersion vanishes.}. This property is crucial for robust physical wave modelling in coastal environments. Indeed, a wave approaching continental shelf undergoes nonlinear transformations: the water depth is decreasing and the wave amplitude grows, which often leads to the formation of undular bores. The model has to follow these transformations. Mathematically it means that the model equations should encompass a range of physical regimes varying from fairly shallow water to intermediate depths \cite{Gobbi2000}. There exists an option of coupling different hydrodynamic models as it was done \eg in \cite{Lovholt2010}. However, the coupling represents a certain number of difficulties, \eg
\begin{itemize}
  \item Boundary conditions at artificial interfaces?
  \item How to determine automatically the physical regime?
  \item Dynamic evolution and handling of model applicability areas\dots
\end{itemize}
Consequently, in this study we let the physical model to do this work for us.
\end{remark}

\subsubsection{Energy conservation}

We would like to raise the question of energy conservation in nonlinear dispersive wave models. The full Euler equations naturally have this property. So, it is a priori natural to require that a good approximation to Euler equations conserves the energy as well \cite{Fedotova2014}. An energy conservation equation can be established for the base model \eqref{eq:base1}, \eqref{eq:base2} for some choices of the variable $\J(\H,\,\ub)$. For instance, the classical SGN model discussed in the following section enjoys this property (it corresponds to the choice $\J\ \equiv\ \vO$). On moving bottoms this property was discussed in \cite{Fedotova2014}. Here we provide only the final result, \ie the total energy equation for SGN model on a general moving bottom\footnote{Of course, this equation becomes a conservation law only when the bottom is static (but not necessarily flat).}:
\begin{equation*}
  (\H\,\E)_t\ +\ \div\Bigl[\,\H\ub\,\Bigl(\E\ +\ \frac{\Pp}{\H}\Bigr)\,\Bigr]\ =\ -\pc\, h_t\,,
\end{equation*}
where the total energy $\E$ is defined as
\begin{equation*}
  \E\ \eqdef\ \half\,\abs{\ub}^2\ +\ \sixth\, \H^{\,2}\,(\div\ub)^2\ +\ \half\,\H\,(\D h)\,(\div\ub)\ +\ \half\,(\D h)^2\ +\ \frac{g}{2}\;(\H\ -\ 2h)\,.
\end{equation*}
For other choices of the closure $\J(\H,\,\ub)$ this question of energy conservation has to be studied separately.

\begin{remark}
Recently, \textsc{Clamond}, \textsc{Dutykh} \& \textsc{Mitsotakis} (2015) \cite{Clamond2015c} proposed a dispersion-improved SGN-type model which enjoys the energy conservation property. The method employed in that study is the variational approach: the preservation of the variational structure is crucial for the preservation of several invariants.
\end{remark}

\subsubsection{Galilean invariance}

The same questions can be raised about the Galilean invariance property as well. This property is of fundamental importance for any mathematical model that provide a physically sound description of water waves (stemming from Classical Mechanics and Classical Physics). Some thoughts and tentative corrections can be found in \cite{Duran2013, Dutykh2011e}. The base model \eqref{eq:base1}, \eqref{eq:base2} is Galilean invariant under reasonable assumptions on the closure velocity vector $\J$.

Galilean invariance principle states that all mechanical laws are the same in any \emph{inertial} frame of reference \cite{Landau1976}. Consequently, the mathematical form of governing equations should be the same as well. It was proposed by Galileo~\textsc{Galilei} in 1632 \cite{Galilei1632}. Consider the horizontal Galilean boost transformation between two inertial frames of reference:
\begin{equation}\label{eq:boost}
  \x^{\,\prime}\ =\ \x\ +\ \Cc\, t\,, \qquad y^{\,\prime}\ =\ y\,, \qquad t^{\,\prime}\ =\ t\,,
\end{equation}
where $\Cc$ is a constant motion speed of the new coordinate system (with primes) relatively to the initial one (without primes). Notice that scalar quantities such as $\H(\x,\,t)$ and $h(\x,\,t)$ remain invariant since they are defined as distances between two points and distances are preserved by the Galilean transformation \eqref{eq:boost}. Let us see how the horizontal velocity variable changes under the Galilean transformation:
\begin{equation*}
  \u(\x,\,y,\,t)\ \eqdef\ \od{\x}{t}\ =\ \od{\x^{\,\prime}}{t}\ -\ \Cc\ \defeq\ \u^{\,\prime}(\x^{\,\prime},\,y^{\,\prime},\,t^{\,\prime})\ -\ \Cc\,.
\end{equation*}
It is not difficult to understand that the same transformation rule applies to $\ub(\x,\,t)$ regardless if it is defined as a trace or depth-averaged velocity:
\begin{equation*}
  \ub\ =\ \ub^{\,\prime}\ -\ \Cc\,.
\end{equation*}
Indeed, the last claim is obvious for the case of the trace operator. Let us check it for the depth-averaging operator:
\begin{multline*}
  \ub(\x,\,t)\ \eqdef\ \frac{1}{\H}\,\int_{-h}^{\;\eps\,\eta} \u\;\ud y\ =\ \frac{1}{\H^{\,\prime}}\,\int_{-h^{\,\prime}}^{\;\eps\,\eta^{\,\prime}} \bigl(\u^{\,\prime}\ -\ \Cc\bigr)\;\ud y^{\,\prime}\ =\\=\ \frac{1}{\H^{\,\prime}}\,\int_{-h^{\,\prime}}^{\;\eps\,\eta^{\,\prime}} \u^{\,\prime}\;\ud y^{\,\prime}\ -\ \Cc\ \defeq\ \ub^{\,\prime}(\x^{\,\prime},\,t^{\,\prime})\ -\ \Cc\,.
\end{multline*}
If the velocity $\ub(\x,\,t)$ is defined in a different way, its transformation rule has to be studied separately. From the definition \eqref{eq:defJ} it follows that the velocity correction $\J$ should remain invariant under the Galilean boost \eqref{eq:boost} (since it is defined as a difference of two velocities):
\begin{equation}\label{eq:rule}
  \J^{\,\prime}\ \equiv\ \J\,.
\end{equation}
In the following we shall assume that the chosen closure $\J(\H,\,\ub)$ satisfy the last transformation rule.

Finally, let us discuss the invariance of the base model \eqref{eq:base1}, \eqref{eq:base2}. Basically, this property follows from the transformation rule \eqref{eq:rule}, from the fact that $\Cc\ =\ \const$ and the following observation\footnote{Let us prove, for example, the first identity:
\begin{equation*}
  \D \H\ \equiv\ \H_t\ +\ (\ub\scal\grad) \H\ =\ \H^{\,\prime}_{t^{\prime}}\ +\ \Cc\,\grad \H^{\,\prime}\ +\ \bigl((\ub^{\,\prime} - \Cc)\scal\grad\bigr) \H^{\,\prime}\ =\ \H^{\,\prime}_{t^{\prime}}\ +\ (\ub^{\,\prime}\scal\grad) \H^{\,\prime}\ \equiv\ \D^{\,\prime} \H^{\,\prime}\,.
\end{equation*}}:
\begin{equation*}
  \D \H\ \equiv\ \D^{\,\prime} \H^{\,\prime}\,, \qquad
  \D\ub\ \equiv\ \D^{\,\prime} \ub^{\,\prime}\,.
\end{equation*}
The pressure variables $\Pp$ and $\pc$ remain invariant as well, since they depend on velocity through $\Rr_1$ and $\Rr_2$, which depend in their term only on the full derivative and divergence of the velocity $\ub$. Thus, the base model \eqref{eq:base1}, \eqref{eq:base2} is Galilean invariant under not very restrictive assumptions made above.

\begin{remark}
Many \textsc{Boussinesq}-type equations derived and published in the literature are not Galilean invariant. As a classical such example we can mention \textsc{Peregrine}'s (1967) system \cite{Peregrine1967}. In \cite{Fedotova2014} it was shown how to derive a weakly nonlinear model from the fully nonlinear one in such a way that the reduced \textsc{Boussinesq}-type model has the Galilean invariance and energy conservation properties.
\end{remark}

\subsection{Serre--Green--Naghdi equations}

The celebrated \textsc{Serre}--\textsc{Green}--\textsc{Naghdi} (SGN) equations can be obtained by choosing the simplest possible closure, \ie
\begin{equation*}
  \J\ \equiv\ \vO\,.
\end{equation*}
This closure follows from the fact that the velocity variable $\ub$ chosen in SGN equations is precisely the depth-averaged velocity. Thus, $\ut(\x,\,t)\ \equiv\ \vO$ and from \eqref{eq:defJ} we have that $\J(\x,\,t)\ \equiv\ \vO$. By substituting the proposed closure into equations \eqref{eq:base1}, \eqref{eq:base2}, we obtain the SGN equations:
\begin{align*}
  \H_t\ +\ \div[\,\H\ub\,]\ &=\ 0\,, \\
  (\H\ub)_t\ +\ \div\Bigl[\,\H\ub\otimesb\ub\ +\ \Pp(\H,\,\ub)\cdot\Id\,\Bigr]\ &=\ \pc\,\grad h\,,
\end{align*}
where $\Pp(\H,\,\ub)$ was defined in \eqref{eq:press2}. The last equation can be written in a non-conservative form as well:
\begin{equation*}
  \ub_t\ +\ (\ub\scal\grad)\ub\ +\ \frac{\grad\Pp}{\H}\ =\ \frac{\pc\,\grad h}{\H}\,.
\end{equation*}
The SGN equations have been rediscovered independently by a number of authors. The steady version of these equations can be already found in \textsc{Rayleigh} (1876) \cite{LordRayleigh1876}. Then, this model in 1D was derived by \textsc{Serre} (1953) \cite{Serre1953, Serre1953a} and by \textsc{Su} \& \textsc{Gardner} (1969) \cite{Su1969}. A modern derivation was done by \textsc{Green}, \textsc{Laws} \& \textsc{Naghdi} (1974) \cite{Green1974}. Later, in Soviet Union this system was derived also by \textsc{Pelinovsky} \& \textsc{Zheleznyak} (1985) \cite{Zheleznyak1985}. More recently, modern derivations of these equations based on variational principles have been proposed. Namely, \textsc{Miles} \& \textsc{Salmon} (1985) \cite{Miles1985} gave a derivation in Lagrangian (\eg particle) description. The variational derivation in Eulerian description was given by \textsc{Fedotova} \& \textsc{Karepova} (1996) \cite{Fedotova1996} and later by \textsc{Kim} \etal (2001) \cite{Kim2001} and \textsc{Clamond} \& \textsc{Dutykh} (2012) \cite{Clamond2009}. Recently the multi-symplectic structure for SGN equations was proposed in \cite{Chhay2016}.


\subsection{Other particular cases}

The scope of the present section is slightly broader than its title may suggest. More precisely, we consider the whole class of models where the velocity variable is defined on a certain surface inside the fluid, see equation \eqref{eq:defY} for the definition. We show in this section that the base model \eqref{eq:base1}, \eqref{eq:base2} can be closed using the partial irrotationality condition. Namely, we assume that only two horizontal components of vorticity vanish, \ie
\begin{equation}\label{eq:assume}
  \u_y\ =\ \mu^2\, \grad v\,.
\end{equation}
Integration of this identity over $y$ and using representations \eqref{eq:hor}, \eqref{eq:vert} leads
\begin{equation*}
  \ut(\x,\,y,\,t)\ =\ -(y + h)\,\bigl[\,\grad(\D h)\ +\ \grad h\,(\div\ub)\,\bigr]\ -\ \frac{(y + h)^2}{2}\grad(\div\ub)\ +\ \left.\ut\right|_{y\, =\, -h}\ +\ \O(\mu^2)\,.
\end{equation*}
Consequently, from \eqref{eq:hor} we obtain
\begin{equation}\label{eq:uz}
  \u(\x,\,y,\,t)\ =\ \ub\ +\ \mu^2\,\Bigl[(y + h)\,\A\ +\ \frac{1}{2}\,(y + h)^2\,\B\ +\ \Cs\Bigr]\ +\ \O(\mu^4)\,,
\end{equation}
where we introduced for simplicity the following notation:
\begin{align*}
  \A(\x,\,t)\ &\eqdef\ -\grad(\D h)\ -\ \grad h\,(\div\ub)\,, \\
  \B(\x,\,t)\ &\eqdef\ -\grad(\div\ub)\,, \\
  \Cs(\x,\,t)\ &\eqdef\ \left.\ut\right|_{y\, =\, -h}\,.
\end{align*}
Let us evaluate both sides of equation \eqref{eq:uz} at $y_\sigma\ =\ \Y_\sigma(\x,\,t)$. According to \eqref{eq:defY} we must have
\begin{equation*}
  \u\bigl(\x,\,\Y_\sigma(\x,\,t),\,t\bigr)\ \equiv\ \ub(\x,\,t)\,.
\end{equation*}
Consequently, we have
\begin{equation*}
  \Cs(\x,\,t)\ \equiv\ -(y_\sigma + h)\,\A\ -\ \frac{1}{2}\,(y_\sigma + h)^2\,\B\,.
\end{equation*}
Thus, coefficient $\Cs$ can be eliminated from \eqref{eq:uz} to give the following representation
\begin{equation*}
  \u(\x,\,y,\,t)\ =\ \ub\ +\ \mu^2\,\Bigl[(y - y_\sigma)\,\A\ +\ \frac{1}{2}\,\bigl[\,(y + h)^2\ -\ (y_\sigma + h)^2\,\bigr]\,\B\Bigr]\ +\ \O(\mu^4)\,.
\end{equation*}
Substituting the last result into equation \eqref{eq:defJ} yields the required closure relation:
\begin{equation}\label{eq:closure1}
  \J(\H,\,\ub)\ =\ \Bigl[\,\frac{\H}{2}\ -\ (y_\sigma + h)\Bigr]\,\A\ +\ \Bigl[\frac{1}{6}\, \H^{\,2}\ -\ \frac{1}{2}\,(y_\sigma + h)^2\,\Bigr]\,\B\ +\ \O(\mu^2)\,.
\end{equation}
To summarize, under the assumption \eqref{eq:assume} that the first two components of the vorticity field vanish, we can propose a closure to the base model, after neglecting the terms of order $\O(\mu^2)$ in \eqref{eq:closure1}.

\bigskip
\paragraph*{Depth-averaged velocity.}

It is interesting to obtain also the 3D velocity reconstruction formula in the case, where $\ub(\x,\,t)$ is defined as the depth-averaged velocity \eqref{eq:deptha}. To do it, we average the equation \eqref{eq:uz} over the depth:
\begin{equation*}
  \frac{1}{\H}\;\int_{-h}^{\,\eps\eta}\u(\x,\,y,\,t)\;\ud y\ =\ \ub(\x,\,t)\ +\ \mu^2\,\Bigl[\,\frac{\H}{2}\;\A\ +\ \frac{\H^{\,2}}{6}\;\B\ +\ \Cs\,\Bigr]\ +\ \O(\mu^2)\,.
\end{equation*}
Using the definition \eqref{eq:deptha} of the depth-averaged velocity, we conclude that
\begin{equation*}
  \Cs\ =\ -\frac{\H}{2}\;\A\ -\ \frac{\H^{\,2}}{6}\;\B\ +\ \O(\mu^2)\,.
\end{equation*}
By substituting the last expression into \eqref{eq:uz} we obtain the desired representation:
\begin{multline}\label{eq:vel}
  \u(\x,\,y,\,t)\ =\ \ub(\x,\,t)\ +\\ \mu^2\,\biggl[\,\Bigl(\frac{\H}{2}\ -\ y\ -\ h\Bigr)\cdot\Bigl(\grad\D h\ +\ (\div\ub)\,\grad h\Bigr)\ +\ \Bigl(\frac{\H^{\,2}}{6}\ -\ \frac{(y + h)^2}{2}\Bigr)\,\grad(\div\ub)\,\biggr]\ +\ \O(\mu^4)\,.
\end{multline}
The last formula will be used in \cite{Khakimzyanov2016} in order to reconstruct the 3D field under a propagating wave, which undergoes some nonlinear transformations. Formula \eqref{eq:vel} shows also that in shallow water flows the velocity distribution in the vertical coordinate $y$ is nearly quadratic.

\begin{remark}
We underline that formula \eqref{eq:vel} is obtained under the assumption that the flow is irrotational. Without this assumption, in the most general case we can only use formula \eqref{eq:hor} by neglecting terms of the order $\O(\mu^2)\,$. In other words, the velocity variable $\ub(\x,\,t)$ approximates the 3D velocity field $\u(\x,\,y,\,t)$ throughout the fluid to the order $\O(\mu^2)\,$. However, in many applications this accuracy is not enough.
\end{remark}


\subsubsection{Lynett--Liu's model}

It can be shown that the base model \eqref{eq:base1}, \eqref{eq:base2} supplemented by the proposed closure \eqref{eq:closure1} is asymptotically equivalent to the well-known \textsc{Lynett}--\textsc{Liu} (2002) model derived in \cite{Lynett2002} under an additional assumption that the initial 3D flow is \emph{irrotational}. This claim is true only up to the approximation order $\O(\mu^4)$ and it can be checked by straightforward but tedious calculations.

Various choices of the level $y_\sigma$, where the horizontal velocity is defined, allow to obtain in a straightforward manner the fully nonlinear analogues of various existing models. Some of popular choices are discussed below.

\subsubsection{Mei--Le M\'ehaut\'e's model}

Consider the horizontal velocity variable defined at the bottom, \ie
\begin{equation*}
  y_\sigma\ =\ -h(\x,\,t)\,.
\end{equation*}
Substituting this value into \eqref{eq:closure1} we obtain straightforwardly the following closure:
\begin{equation}\label{eq:mei}
  \J(\H,\,\ub)\ =\ \frac{1}{2}\, \H\,\A\ +\ \frac{1}{6}\, \H^{\,2}\,\B\ +\ \O(\mu^2)\,.
\end{equation}
In this way, the base model \eqref{eq:base1}, \eqref{eq:base2} with the last closure becomes the celebrated \textsc{Mei}--\textsc{Le M\'ehaut\'e} (1966) model \cite{Mei1966}.

\subsubsection{Peregrine's model and its generalizations}

In 1967 \textsc{Peregrine} \cite{Peregrine1967} considered a weakly nonlinear model with $y_\sigma\ =\ 0$. The fully nonlinear analogue of \textsc{Peregrine}'s model can be obtained if we take
\begin{equation*}
  y_\sigma\ =\ \eps\,\eta\,(\x,\,t)\,.
\end{equation*}
Closure relation \eqref{eq:closure1} then becomes:
\begin{equation*}
  \J(\H,\,\ub)\ =\ -\frac{1}{2}\, \H\,\A\ -\ \frac{1}{3}\, \H^{\,2}\,\B\ +\ \O(\mu^2)\,,
\end{equation*}
and base model \eqref{eq:base1}, \eqref{eq:base2} becomes the fully nonlinear \textsc{Peregrine}'s system. The momentum equation of this model takes a very simple form, when the \textsc{Boussinesq} regime is considered:
\begin{equation*}
  \ub_t\ +\ (\ub\scal\grad)\ub\ +\ \eps\,\grad\eta\ =\ \vO\,.
\end{equation*}
In other words, if initially the vertical component of vorticity is zero, then it is so for all times, \ie
\begin{equation}\label{eq:vort0}
  \ubar_{2,\,x_1}\ -\ \ubar_{1,\,x_2}\ =\ 0, \qquad \forall t\ \geq\ 0\,.
\end{equation}
The last assertion is true only in \textsc{Boussinesq} approximation in for the Cauchy problem. The irrotationality can break when boundary conditions are applied on finite domains, \cite{Dougalis2010}.

\subsubsection{Nwogu's model and its generalizations}
\label{sec:disperse}

In 1993 \textsc{Nwogu} proposed the following choice \cite{Nwogu1993}:
\begin{equation*}
  y_\sigma\ \approx\ -\beta\cdot h(\x,\,t)\,, \qquad \beta\ \approx\ 0.531\,.
\end{equation*}
This choice was motivated by linear dispersion relation considerations (optimization of dispersive characteristics). The nonlinearity of \textsc{Nwogu}'s model was improved in \eg \cite{Kennedy2001, Shi2012}. The idea consists in finding surface between the bottom $y\ =\ -h(\x,\,t)$ and free surface $y\ =\ \eps\,\eta(\x,\,t)$ (instead of the bottom and $y\ =\ 0$ in weakly nonlinear considerations). In this way, a free parameter $\beta\ \in\ [0,\,1]$ at our disposal:
\begin{equation*}
  y_\sigma(\x,\,t)\ =\ -\beta\, h(\x,\,t)\ +\ (1 - \beta)\,\eps\,\eta\,(\x,\,t)\,.
\end{equation*}
In this case the closure relation becomes:
\begin{equation*}
  \J(\H,\,\ub)\ =\ \bigl(\,\beta\ -\ \frac{1}{2}\,\bigr)\, \H\,\A\ -\ \frac{\H^{\,2}}{6}\,\bigl(\,3\beta^2\ -\ 6\beta\ +\ 2\,\bigr)\,\B\ +\ \O(\mu^2)\,.
\end{equation*}
The `optimal' value of $\beta$ will coincide with that given by \textsc{Nwogu} \cite{Nwogu1993} since linearizations of both models coincide.

\subsubsection{Aleshkov's model}

As the last example, we show here how to obtain \textsc{Aleshkov}'s (1996) model \cite{Aleshkov1996}, which was generalized later to include moving bottom effects in \cite{Fedotova2009}. \textsc{Aleshkov}'s model (with moving bottom) can be obtained from the base model \eqref{eq:base1}, \eqref{eq:base2} if we adopt the following closure:
\begin{equation}\label{eq:al}
  \J(\H,\,\ub)\ =\ -(\D h)\,\grad h\ +\ \frac{1}{2}\, \H\,\A\ +\ \frac{1}{6}\, \H^{\,2}\,\B\ +\ \O(\mu^2)\,.
\end{equation}
This closure is similar to \textsc{Mei}--\textsc{Le M\'ehaut\'e} closure \eqref{eq:mei} except for the first term. The horizontal velocity in \textsc{Aleshkov}'s model does not coincide with the horizontal fluid velocity at any surface inside fluid bulk. Instead, \textsc{Aleshkov}'s velocity variable is given by the gradient of the velocity potential evaluated at solid bottom. For non-flat bottoms it does not coincide with $\left.\u\right|_{y\, =\, -h}$. These subtle differences are discussed in some detail in Appendix~\ref{app:diff}. Since this model is not widely known, we give here the governing equations:
\begin{multline}\label{eq:aleshkov1}
  \H_t\ +\ \div\bigl[\,\H\ub\,\bigr]\ =\\ \mu^2\,\div\Bigl[\,\H(\grad h)\,\D h\ +\ \frac{\H^{\,2}}{2}\;\bigl[\,\grad(\D h)\ +\ (\div\ub)\,\grad h\,\bigr]\ +\ \frac{\H^{\,3}}{6}\;\grad(\div\ub)\,\Bigr]\,,
\end{multline}
\begin{equation}\label{eq:aleshkov2}
  \ub_t\ +\ (\ub\scal\grad)\,\ub\ +\ \eps\,\grad\eta\ =\ \mu^2\,\grad\,\Bigl[\,\H\,\Rr_2\ +\ \frac{\H^{\,2}}{2}\;\Rr_1\ +\ \frac{1}{2}\;(\D h)^2\,\Bigr]\ +\ \O(\mu^4)\,.
\end{equation}
One big advantage of equations above is that the irrotational flow is preserved by its dynamics of equations \eqref{eq:aleshkov1}, \eqref{eq:aleshkov2} in the sense of definition given in equation \eqref{eq:vort0}. The proof of this fact is given in Appendix~\ref{app:aleshkov}.


\section{Weakly-nonlinear models}
\label{sec:weak}

We considered the fully nonlinear version of the base model \eqref{eq:base1}, \eqref{eq:base2} previously since the small amplitude assumption was never used (even if we introduced formally the nonlinearity parameter $\eps$). The only constitutive assumption employed was the long wave hypothesis or, in other words, the waves are only weakly dispersive. In the present section we derive a weakly nonlinear variant of the base model \eqref{eq:base1}, \eqref{eq:base2}. In this way we achieve a further simplification of governing equations. Moreover, we shall work in the so-called \textsc{Boussinesq} regime:
\begin{equation}\label{eq:bouss}
  \eps\ =\ \O(\mu^2)\ \quad\ \Longleftrightarrow\ \quad \Su\ =\ \O(1)\,,
\end{equation}
where $\Su\ \eqdef\ \dfrac{\eps}{\mu^2}\ \equiv\ \dfrac{\alpha\,\ell^{\,2}}{d^{\,3}}$ is the so-called \textsc{Stokes}--\textsc{Ursell} number \cite{Ursell1953}. In other words, we assume that the nonlinearity and dispersion parameters have \emph{approximatively} the same order of magnitude. It is under this assumption that one can obtain numerous \textsc{Boussinesq}-type models \cite{BCS, DMII}. Sometimes the simplifying \textsc{Boussinesq} assumption \eqref{eq:bouss} is accompanied also by explicitly (or implicitly) stated assumptions on the bottom variations, \eg $\norm{\grad h}\ \sim\ \O(\eps)\ \simeq\ \O(\mu^2)\,$, as it is the case for the base model.

The most difficult task here is to keep as many good properties of the base model as possible, while simplifying the governing equations. It is not always possible and some illustrations will be given below.


\subsection{Weakly nonlinear base model}

In the present Section we derive the Weakly Nonlinear Base Model (WNBM) starting from the base model equations \eqref{eq:base1}, \eqref{eq:base2}. The first goal here is to preserve at least the conservative form of the equations when simplifying the base model.

First of all, we notice that the vector $\J$ always enters into governing equations with coefficient $\mu^2$, \ie $\mu^2\,\J\,$. Consequently, under the assumption \eqref{eq:bouss}, the vector $\J$ can be formally split as
\begin{equation*}
  \J\ =\ \underbrace{\J_0\ +\ \J_1}_{\O(1)}\ +\ \O(\mu^2)\,,
\end{equation*}
where $\J_0$ contains all the terms independent of the system solution and $\J_1$ contains everything else involving $\eta, \ub$. For instance, to illustrate this idea for the closure relation \eqref{eq:closure1}, which gives the \textsc{Lynett}--\textsc{Liu}'s model, we have the following decomposition:
\begin{align}\label{eq:clo1}
  \J_0\ &=\ \Bigl(\frac{h}{2}\ +\ y_\sigma\Bigr)\,\grad h_t\,, \\
  \J_1\ &=\ \Bigl(\frac{h}{2}\ +\ y_\sigma\Bigr)\cdot\bigl(\grad(\ub\scal\grad h)\ +\ (\div\ub)\grad h\bigr)\ -\ \Bigl(\frac{h^2}{6}\ -\ \frac{(y_\sigma + h)^2}{2}\Bigr)\,\grad(\div\ub)\,. \label{eq:clo2}
\end{align}
From now on we use the following notation for the main part of vector $\J$:
\begin{equation*}
  \J^\flat\ \eqdef\ \J_0\ +\ \J_1\,.
\end{equation*}
The mass conservation equation for WNBM model is directly obtained from \eqref{eq:mass}:
\begin{equation}\label{eq:massW}
  \H_t\ +\ \div(\H\,\ub)\ =\ -\mu^2\,\div(h\,\J^\flat)\,.
\end{equation}
We notice that the last equation is in the conservative form as well. In a similar way, we obtain the weakly nonlinear analogue of the momentum conservation equation:
\begin{equation}\label{eq:momentW}
  (\H\,\ub)_t\ +\ \div(\H\,\ub\otimes\ub)\ +\ \grad\Pp^{\,\flat}\ =\ \pc^{\,\flat}\,\grad h\ -\ \mu^2\Bigl[\,(h\,\J^\flat)_t\ +\ \div\bigl(h\,\J_0\otimes\ub\ +\ h\,\ub\otimes\J_0\bigr)\,\Bigr]\,.
\end{equation}
In some cases, it is useful to have also a non-conservative form of the momentum conservation equation \eqref{eq:momentW}, which can be obtained using the weakly nonlinear form of the mass conservation \eqref{eq:massW}:
\begin{multline}\label{eq:momentW2}
  \ub_t\ +\ (\ub\scal\grad)\ub\ +\ \frac{\grad\Pp^{\,\flat}}{\H}\ =\ \frac{\pc^{\,\flat}\,\grad h}{\H}\\ -\ \frac{\mu^2}{\H}\,\Bigl[\,(h\,\J^\flat)_t\ -\ \ub\,\div(h\,\J^\flat)\ +\ \div\bigl(h\,\J_0\otimes\ub\ +\ h\,\ub\otimes\J_0\bigr)\,\Bigr]\,.
\end{multline}
To complete the description of the WNBM, we have to explain how to compute the non-hydrostatic pressure in this model:
\begin{align*}
  \Pp^\flat\ &\eqdef\ \frac{\H^{\,2}}{2}\ -\ \mu^2\biggl[\,\frac{h^3}{3}\;\Rr_1^{\,\flat}\ +\ \frac{h^2}{2}\;\Rr_2^{\,\flat}\,\biggr]\,, \\
  \pc^\flat\ &\eqdef\ \H\ -\ \mu^2\,\biggl[\,\frac{h^2}{2}\;\Rr_1^{\,\flat}\ +\ h\,\Rr_2^{\,\flat}\,\biggr]\,,
\end{align*}
where
\begin{equation*}
  \Rr_1^{\,\flat}\ \eqdef\ (\div\ub)_t\,, \qquad
  \Rr_2^{\,\flat}\ \eqdef\ h_{tt}\ +\ 2\,\ub\scal\grad h_t\ +\ \ub_t\scal\grad h\,.
\end{equation*}
We underline that the non-conservative form \eqref{eq:momentW2} contains one nonlinear dispersive term $\ub\,\div(h\,\J_1)$ while in the conservative form \eqref{eq:momentW} all dispersive terms are \emph{linear}. Equations \eqref{eq:massW}, \eqref{eq:momentW} constitute the WNBM. Below we derive some important particular cases of WNBM.


\subsubsection{Depth-averaged WNBM}
\label{sec:deptha}

Consider a particular case of the WNBM when the velocity variable is chosen to be depth-averaged. In this case we showed above that $\J\ \equiv\ \vO\,$. Consequently, $\J^\flat\ \equiv\ \vO$ as well. WNBM equations \eqref{eq:massW}, \eqref{eq:momentW} take the simplest form in this particular case:
\begin{align*}
  \H_t\ +\ \div(\H\,\ub)\ &=\ 0\,, \\
  (\H\,\ub)_t\ +\ \div(\H\,\ub\otimes\ub)\ +\ \grad\Pp^{\,\flat}\ &=\ \pc^{\,\flat}\,\grad h\,.
\end{align*}
On the flat bottom the last equation becomes even simpler:
\begin{equation*}
  (\H\,\ub)_t\ +\ \div(\H\,\ub\otimes\ub)\ +\ \grad\Pp^{\,\flat}\ =\ \vO\,.
\end{equation*}
The equivalent non-conservative form of the momentum conservation equation (on uneven bottoms) is
\begin{equation}\label{eq:per}
  \ub_t\ +\ (\ub\scal\grad)\,\ub\ +\ \alpha\,\grad\eta\ =\ \underbrace{\frac{\mu^2}{\H}\;\biggl\{\grad\biggl[\,\frac{h^3}{3}\;\Rr_1^{\,\flat}\ +\ \frac{h^2}{2}\;\Rr_2^{\,\flat}\,\biggr]\ -\ \biggl[\,\frac{h^2}{2}\;\Rr_1^{\,\flat}\ +\ h\,\Rr_2^{\,\flat}\,\biggr]\,\grad h\biggr\}}_{(\mathlarger{\checkmark})}\,.
\end{equation}


\subsubsection{Peregrine's system}
\label{sec:per}

In the pioneering work \cite{Peregrine1967} \textsc{Peregrine} derived a weakly-nonlinear model over a \emph{stationary bottom}, \ie $h_t\ \equiv\ 0\,$. The mass conservation in \textsc{Peregrine}'s system coincides exactly with the mass conservation equation from the previous Section~\ref{sec:deptha}. We show below that the \textsc{Peregrine}'s momentum conservation can be obtained from the non-conservative equation \eqref{eq:per} under the \textsc{Boussinesq} assumption \eqref{eq:bouss}. The right-hand side $(\checkmark)$ can be rewritten as
\begin{multline*}
  \frac{(\checkmark)}{\mu^2}\ =\ \frac{1}{h}\;\grad\biggl[\,\frac{h^3}{3}\;(\div\ub)_t\ +\ \frac{h^2}{2}\;(\ub_t\scal\grad h)\,\biggr]\ -\ \biggl[\,\frac{h}{2}\;(\div\ub)_t\ +\ \ub_t\scal\grad h\,\biggr]\;\grad h\ =\\ =\ \biggl[\,\frac{h}{2}(\div\ub)\,\grad h\ +\ \frac{h^2}{3}\;\grad(\div\ub)\ +\ \frac{h}{2}\;\grad(\ub\scal\grad h)\,\biggr]_t\,.
\end{multline*}
Then, we use the relation
\begin{equation}\label{eq:rel}
  \ub\scal\grad h\ =\ \div(h\,\ub)\ -\ h\,\div\ub\,.
\end{equation}
Finally, we obtain the right-hand side of \textsc{Peregrine}'s model:
\begin{equation*}
  \frac{(\checkmark)}{\mu^2}\ =\ \biggl[\,\frac{h}{2}\;\grad\bigl(\div(h\,\ub)\bigr)\ -\ \frac{h^2}{6}\;\grad(\div\ub)\,\biggr]_t\,.
\end{equation*}
Hence, the non-conservative momentum equation reads
\begin{equation*}
  \ub_t\ +\ (\ub\scal\grad)\,\ub\ +\ \alpha\,\grad\eta\ =\ \underbrace{\mu^2\,\biggl[\,\frac{h}{2}\;\grad\bigl(\div(h\,\ub)\bigr)\ -\ \frac{h^2}{6}\;\grad(\div\ub)\,\biggr]_t}_{\simeq\ (\mathlarger{\checkmark})}\,.
\end{equation*}
However, the simplifications we made above were drastic in some sense. For instance, the \textsc{Peregrine}'s model cannot be recast in a conservative form even on a flat bottom. It goes without saying that the energy equation cannot be established for this model either. These are the main drawbacks of the weakly nonlinear \textsc{Peregrine}'s system. Moreover, the numerical schemes based on non-conservative equations may be divergent \cite{LeVeque1992}. Despite all this critics, the \textsc{Peregrine}'s system supplemented with moving bottom effects (\ie $h_t\ \neq\ 0$) was successfully used to model wave generation in closed basins \cite{Dutykh2011d, Nersisyan2012}.

It is interesting to note that the depth-averaged WNBM and \textsc{Peregrine}'s system give the same linearisation over the flat bottom $h(x)\ \equiv\ d$:
\begin{align*}
  \eta_t\ +\ d\,\div\ub\ &=\ 0\,, \\
  \ub_t\ +\ \alpha\,\grad\eta\ &=\ \mu^2\,\frac{d^{\,2}}{3}\;\grad(\div\ub_t)\,.
\end{align*}
In particular, it implies that dispersive properties are the same.


\subsubsection{WNBM with the velocity given on a surface}

When the velocity variable is defined on a surface in the fluid bulk as in \eqref{eq:defY}, WNBM equations are \eqref{eq:massW}, \eqref{eq:momentW} and the closure relation for variable $\J^{\,\flat}$ is given by formulas \eqref{eq:clo1}, \eqref{eq:clo2}. Consequently, the dispersive terms are present in both mass and momentum conservation equations. Moreover, in the case of the stationary bottom ($h_t\ \equiv\ 0$) we have automatically that $\J_0\ \equiv\ \vO\,$. Consequently, the WNBM equations with this choice of the velocity variable read:
\begin{align*}
  \H_t\ +\ \div(\H\,\ub)\ &=\ -\mu^2\,\div(h\,\J_1)\,, \\
  (\H\,\ub)_t\ +\ \div(\H\,\ub\otimes\ub)\ +\ \grad\Pp^{\,\flat}\ &=\ \pc^{\,\flat}\,\grad h\ -\ \mu^2\,(h\,\J_1)_t\,.
\end{align*}
The last equation can be recast in the non-conservative form:
\begin{multline*}
  \ub_t\ +\ (\ub\scal\grad)\,\ub\ +\ \alpha\,\grad\eta\ =\\ 
  \underbrace{\frac{\mu^2}{\H}\;\biggl\{\grad\biggl[\,\frac{h^3}{3}\;\Rr_1^{\,\flat}\ +\ \frac{h^2}{2}\;\Rr_2^{\,\flat}\,\biggr]\ -\ \biggl[\,\frac{h^2}{2}\;\Rr_1^{\,\flat}\ +\ h\,\Rr_2^{\,\flat}\,\biggr]\,\grad h\ -\ (h\,\J_1)_{\,t}\ +\ \ub\,\div(h\,\J_1)\biggr\}}_{(\mathlarger{\eth})}\,.
\end{multline*}
Below we show an important application of this variant of the WNBM.


\subsubsection{Nwogu's system}

\textsc{Nwogu}'s model was derived in \cite{Nwogu1993} under the assumption of the stationary bottom ($h_t\ \equiv\ 0$) that we adopt here as well. First of all, the expression \eqref{eq:clo2} can be transformed using the relation \eqref{eq:rel}:
\begin{equation*}
  \J_1\ =\ \Bigl(\frac{h}{2}\ +\ y_\sigma\Bigr)\,\grad\bigl(\div(h\,\ub)\bigr)\ +\ \Bigl(\frac{y_\sigma^2}{2}\ -\ \frac{h^2}{2}\Bigr)\,\grad(\div\ub)\,.
\end{equation*}
In this way we obtain straightforwardly the mass conservation equation of \textsc{Nwogu}'s system \cite{Nwogu1993}. In order to obtain the momentum equation of \textsc{Nwogu}'s system, first we neglect in $(\eth)$ the nonlinear dispersive term $\ub\,\div(h\,\J_1)\,$. Then, the non-hydrostatic pressure terms are transformed similarly to \textsc{Peregrine}'s system case studied above in Section~\ref{sec:per}. So, the right-hand side $(\eth)$ of WNBM becomes:
\begin{multline*}
  \frac{(\eth)}{\mu^2}\ =\ \frac{h}{2}\;\grad\bigl(\div(h\,\ub)_t\bigr)\ -\ \frac{h^2}{6}\;\grad(\div\ub)_t\\ -\ \Bigl(\frac{h}{2}\ +\ y_\sigma\Bigr)\,\grad\bigl(\div(h\,\ub)_t\bigr)\ -\ \Bigl(\frac{y_\sigma^2}{2}\ -\ \frac{h^2}{6}\Bigr)\,\grad(\div\ub)_t\\ \equiv\ -\biggl[\,y_\sigma\,\grad\bigl(\div(h\,\ub)_t\bigr)\ +\ \frac{y_\sigma^2}{2}\;\grad(\div\ub)_t\,\biggr]\,.
\end{multline*}
As a result, we obtain the momentum equation of \textsc{Nwogu}'s system \cite{Nwogu1993}:
\begin{equation*}
  \ub_t\ +\ (\ub\scal\grad)\,\ub\ +\ \alpha\,\grad\eta\ =\ -\underbrace{\mu^2\,\biggl[\,y_\sigma\,\grad\bigl(\div(h\,\ub)_t\bigr)\ +\ \frac{y_\sigma^2}{2}\;\grad(\div\ub)_t\,\biggr]}_{\simeq\ (\mathlarger{\eth})}\,.
\end{equation*}
Using the low-order linear terms in the dispersive terms again, other asymptotic equivalent models can also be derived, \cite{Mitsotakis2007}.
 
The WNBM equations and \textsc{Nwogu}'s system linearize on the flat bottom $h(x)\ \equiv\ d$ to the same equations:
\begin{align*}
  \eta_t\ +\ d\,\div\ub\ &=\ -\mu^2\,d^{\,3}\Bigl(\beta\ +\ \frac{1}{3}\Bigr)\,\div\bigl(\grad(\div\ub)\bigr)\,, \\
  \ub_t\ +\ \alpha\,\grad\eta\ &=\ -\mu^2\,d^{\,2}\,\beta\,\grad\bigl(\div\ub_t\bigr)\,,
\end{align*}
where we introduced the following parameter:
\begin{equation*}
  \beta\ \eqdef\ \frac{y_\sigma}{d}\ +\ \frac{y_\sigma^2}{2\,d^{\,2}}\,.
\end{equation*}
However, in the nonlinear case the WNBM system has the advantage of admitting the conservative form on general (unsteady and uneven) bottoms. This fact can be used to develop efficient numerical algorithms to solve nonlinear dispersive equations numerically. For instance, this conservative property will be exploited in \cite{Khakimzyanov2016} in order to construct adaptive and efficient numerical discretizations.


\section{Discussion}
\label{sec:concl}

We presented a certain number of developments going from the derivation of the base model \eqref{eq:base1}, \eqref{eq:base2} to obtaining some particular models as particular cases. The main conclusions and perspectives of this study are outlined below.

\subsection{Conclusions}

In the present manuscript we attempted to meet two main goals. First of all, we tried to make a review of the continuously growing field of long wave modelling. In particular, we focused on nonlinear dispersive wave models such as some improved \textsc{Boussinesq}-type and \textsc{Serre}--\textsc{Green}--\textsc{Naghdi} (SGN) equations \cite{Serre1953, Green1974, Green1976}, which were not covered in previously published review papers. We apologize in advance if we forgot to mention somebody's contribution to this field. The topic being so broad that it is practically impossible to referred to all the published literature.

Then, we attempted to present a unified approach which incorporates some well-known and some less known models in the same modelling framework. The derivation procedure is based on the minimal set of assumptions. Various models can be obtained as particular cases of the so-called \emph{base model} presented in our study. In the same time, the base model allows to obtain fully nonlinear analogues of previously derived weakly-nonlinear models. The linearizations of old and new models will coincide exactly, hence leaving dispersive characteristics unchanged. Moreover, the resulting models admit an elegant conservative form by construction. The improvement of dispersive characteristics can be achieved by a judicious choice of the closure relation $\J(\H,\,\ub)$ as it was illustrated, for example, in Section~\ref{sec:disperse}.

\subsection{Perspectives}

In the present study we discussed modeling and derivation of models for shallow water waves flowing over uneven bottoms, but the whole system was defined on a flat domain $\Omega$ of the Euclidean space $\R^d$, with dimension $d = 1,\,2$. The bottom represents only a deformation (not necessarily small) of the mean water depth. Among the main perspectives of this study we would like to mention the derivation of fully nonlinear shallow water models defined on more general geometries. In particular, the spherical geometry represents a lot of interest in view of applications to atmospheric sciences. The first steps in this direction have already been made in \cite{Fedotova2011, Fedotova2014a}. The derivation of shallow water equations on a sphere will be discussed in Part~\textsc{III} \cite{Khakimzyanov2016a}.

The numerical discretization of the derived above equations on moving adaptive grids will be considered in details in the companion paper \cite{Khakimzyanov2016} (Part~\textsc{II}), while the numerical simulation of shallow water waves on a sphere will be considered in Part~\textsc{IV} of this series of papers \cite{Khakimzyanov2016b}.


\subsection*{Acknowledgments}
\addcontentsline{toc}{subsection}{Acknowledgments}

This research was supported by RSCF project No 14--17--00219. D. Mitsotakis was supported by the Marsden Fund administered by the Royal Society of New Zealand.


\appendix
\section{Aleshkov's model vs. Mei--Le M\'ehaut\'e's model}
\label{app:diff}

In this Appendix we assume the flow to be irrotational. Consider the fluid velocity potential expansion around the bottom:
\begin{equation}\label{eq:pot}
  \phi(\x,\,y,\,t)\ =\ \phic\ -\ \mu^2(y + h)\bigl(h_t\ +\ \grad\phic\scal\grad h\bigr)\ -\ \mu^2\,\frac{(y + h)^2}{2}\;\grad^2\phic\ +\ \O(\mu^4)\,,
\end{equation}
where $\phic$ is the velocity potential trace at the bottom, \ie
\begin{equation*}
  \phic(\x,\,t)\ \eqdef\ \left.\phi(\x,\,y,\,t)\right|_{y\, =\, -h}\,.
\end{equation*}
A similar formula can be found in \cite{Zheleznyak1985} for the stationary bottom and in \cite{Fedotova2009} for moving bottoms. The horizontal fluid velocity can be readily obtained by differentiating equation \eqref{eq:pot}:
\begin{multline*}
  \u(\x,\,y,\,t)\ \equiv\ \grad\phi\ =\ \grad\phic\ -\ \mu^2\bigl(h_t\ +\ \grad\phic\scal\grad h\bigr)\,\grad h\ -\ \mu^2(y + h)\,\grad\bigl(h_t\ +\ \grad\phic\scal\grad h\bigr)\\ -\ \mu^2(y + h)\,(\grad^2\phic)\,\grad h\ -\ \mu^2\,\frac{(y + h)^2}{2}\;\grad(\grad^2\phic)\ +\ \O(\mu^4)\,.
\end{multline*}
Then, the whole family of models can be obtained by choosing the velocity variable $\ub(\x,\,t)$ at different levels in the fluid. Here we take the velocity at solid bottom:
\begin{equation*}
  \ub(\x,\,t)\ \eqdef\ \left.\u(\x,\,y,\,t)\right|_{y\,=\,-h}\ =\ \grad\phic\ -\ \mu^2\bigl(h_t\ +\ \grad\phic\scal\grad h\bigr)\,\grad h\,.
\end{equation*}
Hence, from definition \eqref{eq:hor} we can compute the expression for $\ut$:
\begin{equation*}
  \ut(\x,\,t)\ =\ -(y + h)\,\grad\bigl(h_t\ +\ \grad\phic\scal\grad h\bigr)\ -\ (y + h)\,(\grad^2\phic)\,\grad h\ -\ \frac{(y + h)^2}{2}\;\grad(\grad^2\phic)\ +\ \O(\mu^2)\,,
\end{equation*}
and taking into account the fact that $\ub\ =\ \grad\phic\ +\ \O(\mu^2)$ we have
\begin{multline*}
  \ut(\x,\,t)\ =\ -(y + h)\,\grad\D h\ -\ (y + h)\,(\div\ub)\,\grad h\ -\ \frac{(y + h)^2}{2}\;\grad(\div\ub)\ +\ \O(\mu^2)\\ 
  \equiv\ (y + h)\,\A\ +\ \frac{(y + h)^2}{2}\; \B\ +\ \O(\mu^2)\,.
\end{multline*}
After applying the depth-averaging operator we obtain the corresponding closure variable:
\begin{equation*}
  \J(\x,\,t)\ \eqdef\ \frac{1}{H}\,\int_{-h}^{\,\eps\,\eta}\ut(\x,\,y,\,t)\;\ud y\ =\ \frac{H}{2}\;\A\ +\ \frac{H^2}{6}\;\B\ +\ \O(\mu^2)\,.
\end{equation*}
It coincides exactly with the closure relation \eqref{eq:mei} given above. This concludes our clarifications regarding \textsc{Mei}--\textsc{Le M\'ehaut\'e}'s model \cite{Mei1966}.

In \textsc{Aleshkov}'s model the velocity variable $\ub(\x,\,t)$ is defined in a different way:
\begin{equation*}
  \ub(\x,\,t)\ \eqdef\ \grad\phic(\x,\,t)\,.
\end{equation*}
Then, the fluid horizontal velocity takes the form
\begin{multline*}
  \u(\x,\,y,\,t)\ =\ \ub\ -\ \mu^2\bigl(h_t\ +\ \grad\phic\scal\grad h\bigr)\,\grad h\ -\ \mu^2(y + h)\,\grad\bigl(h_t\ +\ \grad\phic\scal\grad h\bigr)\\ -\ \mu^2(y + h)\,(\grad^2\phic)\,\grad h\ -\ \mu^2\,\frac{(y + h)^2}{2}\;\grad(\grad^2\phic)\ +\ \O(\mu^4)\ =\\
  \ub\ -\ \mu^2\bigl(h_t\ +\ \ub\scal\grad h\bigr)\,\grad h\ -\ \mu^2(y + h)\,\grad\bigl(h_t\ +\ \ub\scal\grad h\bigr)\\ -\ \mu^2(y + h)\,(\div\ub)\,\grad h\ -\ \mu^2\,\frac{(y + h)^2}{2}\;\grad(\div\ub)\ +\ \O(\mu^4)\ = \\
  \ub\ +\ \mu^2\Bigl[\,-\D h(\grad h)\ +\ (y + h)\,\A\ +\ \frac{(y + h)^2}{2}\,\B\,\Bigr]\ +\ \O(\mu^4)\,.
\end{multline*}
From the last formula it is straightforward to obtain the closure relation \eqref{eq:al} which yields \textsc{Aleshkov}'s model \cite{Aleshkov1996}. It explains also the differences between between \textsc{Aleshkov}'s and \textsc{Mei}--\textsc{Le M\'ehaut\'e}'s models.


\section{Vorticity in Aleshkov's model}
\label{app:aleshkov}

In this Appendix we study how the vertical component of vorticity evolves under the dynamics of \textsc{Aleshkov}'s model \eqref{eq:aleshkov1}, \eqref{eq:aleshkov2}. Consequently, we rewrite equations \eqref{eq:aleshkov2} in the following equivalent form:
\begin{align*}
  \ubar_{1,\,t}\ +\ \ubar_{1}\,\ubar_{1,\,x_1}\ +\ \ubar_{2}\,\ubar_{1,\,x_2}\ +\ \Ru_{x_1}\ &=\ 0\,, \\
  \ubar_{2,\,t}\ +\ \ubar_{1}\,\ubar_{2,\,x_1}\ +\ \ubar_{2}\,\ubar_{2,\,x_2}\ +\ \Ru_{x_2}\ &=\ 0\,,
\end{align*}
where $\Ru$ is a scalar function defined as
\begin{equation*}
  \Ru\ \eqdef\ \eps\,\eta\ -\ \mu^2\,\Bigl[\,H\Rr_2\ +\ \half\,H^2\,\Rr_1\ +\ \half (\D h)^2\,\Bigr]\,.
\end{equation*}
The same equations \eqref{eq:aleshkov2} can be rewritten also as
\begin{align*}
  \ubar_{1\,t}\ -\ \ubar_{2}\,\omega\ +\ \Bigl[\,\Ru\ +\ \frac{\ubar_{1}^{\,2}\ +\ \ubar_{2}^{\,2}}{2}\,\Bigr]_{x_1}\ &=\ 0\,, \\
  \ubar_{2\,t}\ +\ \ubar_{2}\,\omega\ +\ \Bigl[\,\Ru\ +\ \frac{\ubar_{1}^{\,2}\ +\ \ubar_{2}^{\,2}}{2}\,\Bigr]_{x_2}\ &=\ 0\,,
\end{align*}
where we introduced the vertical vorticity function $\omega\ \eqdef\ \ubar_{2,\,x_1}\ -\ \ubar_{1,\,x_2}$. Making a cross differentiation of two last equations and subtracting them yields the following vorticity equation:
\begin{equation}\label{eq:transpoort}
  \omega_t\ +\ [\,\omega\,\ubar_1\,]_{x_1}\ +\ [\,\omega\,\ubar_2\,]_{x_2}\ =\ 0\,.
\end{equation}
Let us assume that initially we have $\omega(\x,\,0)\ \equiv\ 0$ and equation \eqref{eq:transpoort} admits a unique solution. By noticing that $\omega(\x,\,t)\ \equiv\ 0$ solves equation \eqref{eq:transpoort} and satisfies the initial condition, we obtain the required result.

There is a much shorter (but less insightful) proof of the same result. Namely, by definition of the velocity variable $\ub$ in \textsc{Aleshkov}'s model we have:
\begin{equation*}
  \ubar_1\ =\ \phic_{x_1}\,, \qquad \ubar_2\ =\ \phic_{x_2}\,.
\end{equation*}
Then straightforwardly we have
\begin{equation*}
  \omega\ =\ \ubar_{2,\,x_1}\ -\ \ubar_{1,\,x_2}\ =\ (\phic_{x_2})_{x_1}\ -\ (\phic_{x_1})_{x_2}\ \equiv\ 0\,,
\end{equation*}
provided that the trace of the velocity potential at the bottom $\phic$ is a continuously differentiable function.


\section{Acronyms}

In the text above the reader could encounter the following acronyms:

\begin{description}
  \item[BBM] \textsc{Benjamin--Bona--Mahony}
  \item[NSW] Nonlinear Shallow Water
  \item[RLW] Regularized Long Wave
  \item[SGN] \textsc{Serre--Green--Naghdi}
  \item[WNBM] Weakly Nonlinear Base Model
\end{description}


\bigskip\bigskip
\addcontentsline{toc}{section}{References}
\bibliographystyle{abbrv}

\bigskip\bigskip

\end{document}